\documentclass[%
showkeys,reprint,
superscriptaddress,
bibnotes,
 amsmath,amssymb,
 aps,
floatfix,
]{revtex4-1}

\usepackage{graphicx}
\usepackage{dcolumn}
\usepackage{bm}
\usepackage{color}
\usepackage{array,mathtools,amssymb,booktabs}
\usepackage{mhchem}
\newcommand{\lk}{\left(}    \newcommand{\rk}{\right)}

\newcommand{\lK}{\left[} \newcommand{\rK}{\right]}
\begin{document}

\preprint{AIP/123-QED}

\title[]{Optimality in kinetic proofreading and early T-cell recognition: revisiting the speed, energy, accuracy trade-off}

\author{Wenping Cui}
\affiliation{Department of Physics, Boston University, 590 Commonwealth Avenue, Boston, MA 02139}
\affiliation{ Department of Physics, Boston College, 140 Commonwealth Ave, Chestnut Hill, MA 02467}
\author{Pankaj Mehta}%
\email{pankajm@bu.edu}
\affiliation{Department of Physics, Boston University, 590 Commonwealth Avenue, Boston, MA 02139}
\date{\today}

\begin{abstract}
 In the immune system, T cells can quickly discriminate between foreign and self ligands with high accuracy. There is evidence that T-cells achieve this remarkable performance utilizing  a network architecture based on a generalization of kinetic proofreading (KPR). KPR-based mechanisms actively consume energy to increase the specificity beyond what is possible in equilibrium.An important theoretical question that arises is to understand the trade-offs and fundamental limits on accuracy, speed, and dissipation (energy consumption) in KPR and its generalization.  Here, we revisit this question through numerical simulations where we simultaneously measure  the speed, accuracy, and energy consumption of the KPR and adaptive sorting networks for different parameter choices. Our simulations highlight the existence of a ``feasible operating regime'' in the speed-energy-accuracy plane where T-cells can quickly differentiate between foreign and self ligands at reasonable energy expenditure. We give general arguments for why we expect this feasible operating regime to be a generic property of all KPR-based biochemical networks and discuss implications for our understanding of the T cell receptor circuit.
\end{abstract}

\maketitle
%
%
%
\section*{Introduction}
\begin{figure}
\centering
\includegraphics[width=0.5\textwidth]{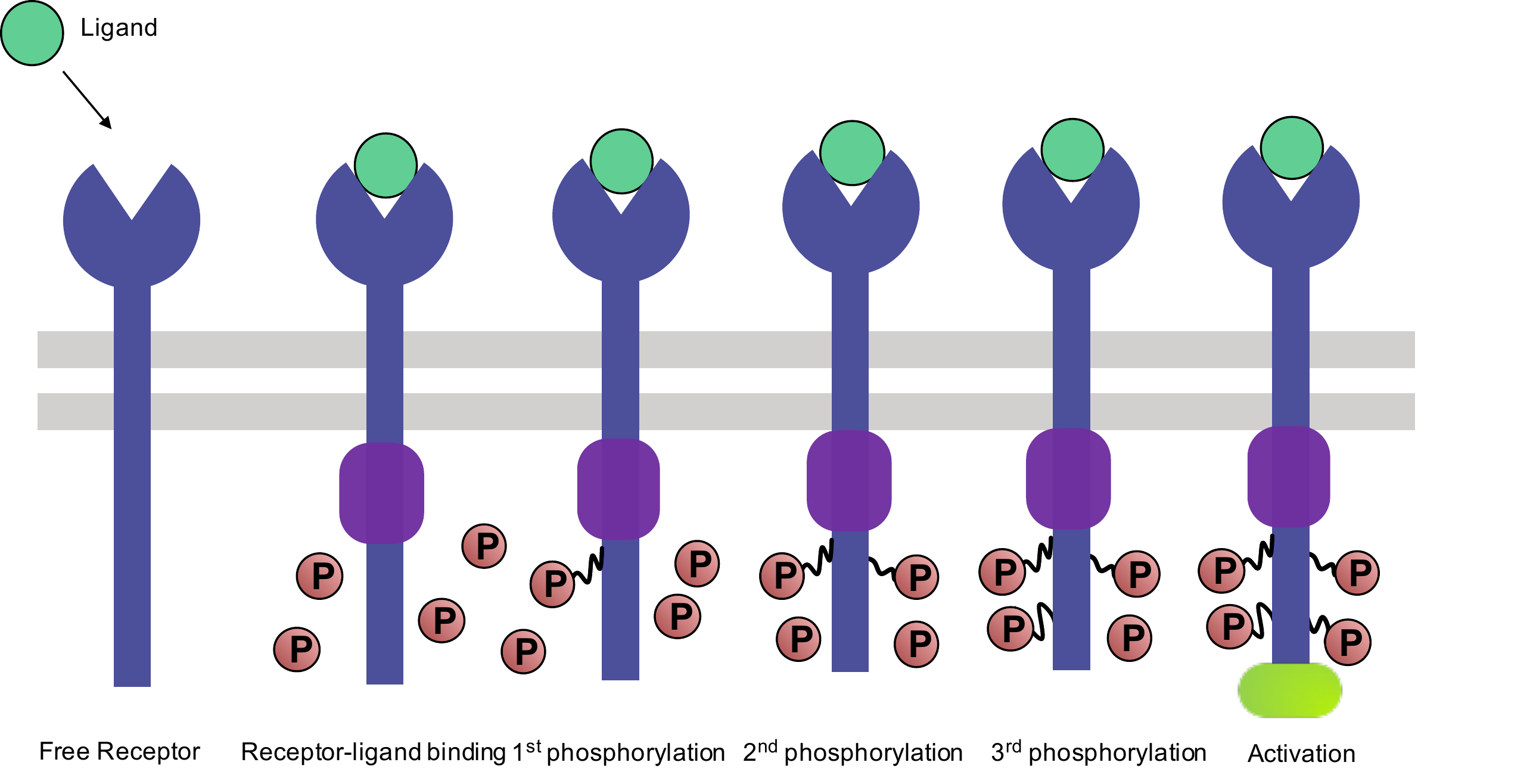}
\caption{ {\bf An overview of T cell activation}. A T-cell is activated when a ligand-binds a TCR receptor long enough to be phosphorylated $n=4$ times. In the ``life-time dogma'' picture 
considered in this paper, foreign ligands bind more strongly to the receptor than self ligands with average disassociation times of foreign and self ligands
given by $\tau_f=10s$ and $\tau_s=1s$ respectively. If the ligand disassociates from the receptor, the receptor is dephosphorlyated and the whole process must begin anew.}
\label{fig1}
\end{figure}

A central problem in immunology is the recognition of
foreign ligands by the immune system. This process is
carried out by specialized immune cells called T-cells which
activate the immune response in the presence of foreign ligands.  Foreign
ligands are presented to T-cells by specialized Antigen Presenting Cells (APCs) that bind a repertoire of self and foreign peptides.
As shown in Fig. \ref{fig1}, T-cells activation occurs when specialized receptors on the
surface of T-cells, called T-cell receptors (TCRs), bind APCs, and activate downstream the TCR signaling network,
leading to an immune response. 

It has been shown that T-cells have a high sensitivity to foreign ligands. A few foreign ligands (less than 10) appearing on the membrane of a T-cell are able to trigger  the immune response\cite{irvine2002direct,chakraborty2014insights}.  Moreover, this decision is made extremely quickly: it only takes 1-5 mins to make the decision to activate or not \cite{stoll2002dynamic}.  Despite the speed with which the response is mounted, T-cells can accurately sense the existence of foreign ligands with an error rate as small as  $10^{-4}-10^{-6}$\cite{mckeithan1995kinetic,alon2006introduction}. This raises natural questions about how the T-cell signaling network can operate with such high speed, sensitivity, and  accuracy. 

Experimental evidence suggests that T-cell activation is set by the binding time of the antigen-receptor complex \cite{feinerman2008quantitative,franccois2016case}.  If the binding time of the ligand to the receptor is below a sharp threshold (3-5 sec), T-cells do not activate. However if the binding time is above this threshold, T-cells activate with extreme sensitivity. This so called `life-time' dogma places stringent conditions on the machinery of the immune response\cite{feinerman2008quantitative}. A lot is known about the biochemical networks that implement this thresholding procedure. The receptor-ligand complexes go through multiple rounds of phosphorylation (throughout we denote the number of phosphorylations by $n$). Within the life-time dogma, an immune response is triggered if the concentration of the ligand-receptor complex that has been phosphorylated $n$ times exceeds a threshold concentration.

The ability of T-cells to discriminate between foreign and self ligands arises from the difference in the binding times of foreign ($\tau_f$) and self ($\tau_s$) ligands\cite{gascoigne2001t}.  Typically, in the immune system, $\tau_s \sim 1s$ and $\tau_f\sim10s$. In equilibrium, this binding time difference cannot account for the incredible accuracy of the T-cell immune response. { Detailed balance places constraints on the chemical reaction rates and the reliability of the discrimination process is ultimately limited by equilibrium thermodynamics\cite{sartori2015thermodynamics}.} This binding time difference can be directly translated into a difference in binding free energies of foreign and self ligands \cite{hopfield1974kinetic, ninio1975kinetic}. Thus, a  biochemical network that works at equilibrium can achieve a minimum error rate of $\tau_s/\tau_f\sim 0.1$, nearly three orders of magnitude smaller than that seen in experiments. 

It is known the immune system can beat this bound by working out-of-equilibrium and consuming energy\cite{mckeithan1995kinetic}. It is now thought that the T-cell employs a form of kinetic proofreading(KPR), first proposed by Hopfield\cite{hopfield1974kinetic} and Ninio\cite{ninio1975kinetic}.  But current understanding of KPR and its implications for immune response have several weaknesses: firstly,  many older theoretical treatments of KPR in the context of T-cell activation involve approximating certain reactions as irreversible making it difficult to consistently calculate energy consumption ; second, it is extremely hard for KPR-based schemes to simultaneously distinguish ligands with similar binding times and operate over a large dynamic range of ligand concentrations.The later shortcoming has been addressed by a generalization of KPR called ``adaptive sorting". In adaptive sorting, an additional feedback couples the KPR cascades in the T-cell through a common kinase that regulates all the phosphorylation of all T-cell receptors \cite{lever2014phenotypic, franccois2008case,franccois2016case,lalanne2013principles,franccois2013phenotypic}. 


A fundamental issue in the study of T-cell activation is to understand the trade-off between different functionalities -- accuracy, speed and dissipation -- in the immune discrimination process. Many works have studied the relation between accuracy and dissipation or accuracy and speed for some KPR-based biochemical network\cite{savageau1979energy,ehrenberg1980thermodynamic,freter1980proofreading,qian2006reducing,murugan2012speed,mehta2016landauer,banerjee102608,das2016limiting,hartich2015nonequilibrium}. Some others have discussed general error rate bounds under power constraints in the context of thermodynamics or information theory\cite{landauer1961irreversibility,bialek2005physical,mora2015physical,lang2014thermodynamics,laughlin2001energy, qian2003thermodynamic,bennett1979dissipation, andrieux2008nonequilibrium, lan2012energy}.   

Early theoretical work suggests that it is always possible to reduce the error of KPR-based mechanisms by waiting longer and/or consuming more energy \cite{savageau1979energy,murugan2012speed}.   However, recent research shows the trade-off between accuracy and speed is not always observed \cite{sartori2013kinetic}.  A recent works which studied KPR in the context of copying polymers and DNA translation and compared experiments with theoretical calculations showed that these systems seem to optimize speed while only suffering minimal costs in accuracy \cite{banerjee102608}. This suggests that even in the context of immune recognition, these trade-offs might not be as stringent as believed and it is worth thoroughly re-examining these tradeoffs in the context of TCR-based circuits.

\begin{figure}
\centering
\includegraphics[width=0.45\textwidth]{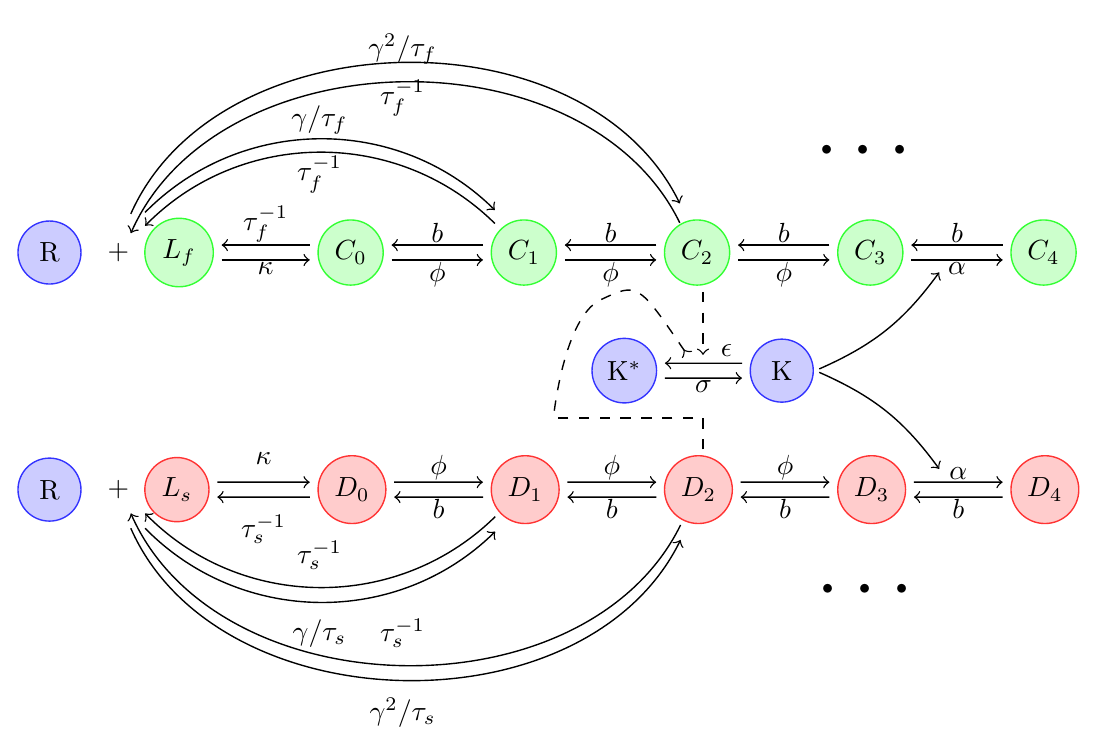}
\caption{{\bf Overview of  KPR-based immune recognition circuits}. Receptors can form complexes with foreign ligands $C$ and self ligands $D$. These complexes disassociate at different rates given by $\tau_f^{-1}$ and $\tau_s^{-1}$, respectively. Once a receptor-ligand complex is formed, it is phosphorylated at a rate $\phi$ and dephosphorylated as a rate $b$. Importantly, ligands can directly form a complex at the $n$-th step of the KPR cascade at a rate $\gamma^n/\tau_i$($i=s, f$). In adaptive sorting circuit, the phosphorlyation/dephosphorylation rates can be modulated by a kinase-dependent feedback loop (see main text).  A full definition of symbols can be found at Table. \ref{table1}. }
\label{fig2}
\end{figure}

In this paper, we calculate the speed, power dissipation, error rate and output signal (the combined concentration of $D_N$ and $C_N$) explicitly for a KPR-based biochemical network for T-cell recognition with and without a feedback that implements adaptive sorting (shown in Fig. \ref{fig2}). We ask if there is a feasible operating region for T-cell activation networks where T cells can make fast and accurate decisions while utilizing energy efficiently.  We find that such a feasible operating region exists for KPR and its generalizations. In the feasible operating region, the response time and power dissipation are consistent with those observed in experiments, implying that many mechanisms of early T-cell recognition are well described by KPR-based models.

\section*{Model}
We start from the adaptive sorting model shown in Fig. \ref{fig2} \cite{franccois2008case,franccois2013phenotypic,franccois2016case,lalanne2013principles}. The receptor, $R$,  can bind a  foreign or self ligand, to form a complex $C_0$ and $D_0$
respectively. This complex can be phosphorylated a maximum of $N$ times. We denote a receptor-ligand complex that has been phosphorlyated $n$ times by $X_n$ with $X=C$ for foreign ligands and $X=D$ for self ligands. The dynamics of the biochemical network can be written as:
\begin{eqnarray}\label{eq1}
\dot{X_0}&=&\kappa RL_i -\left(\tau_i ^{-1}+\phi \right) X_0+b X_1\nonumber\\
\dot{X_n}&=&\!\!\gamma^nRL_i /\tau_i\! \!+\!\!\phi X_{n-1}\!-\! (\phi\! +\!\tau_i ^{-1}\!+\!b)X_{n}\!+\!bX_{n+1}\label{eq22:master}\\
\dot{X_N}&=&\gamma^N RL_i /\tau_i+\alpha K X_{N-1}-(b+\tau_i ^{-1})X_N\nonumber\\
\dot{K}&=&-\epsilon K(C_m+D_m)+\sigma (K_T-K)\nonumber
\end{eqnarray}
where $N>n>0$,  $X\in\{ C,D \}$, and $i \in \{ f,s \}$. $R=R_T-\sum^N_{j=0}(C_j+D_j)$ and $L_i=L^T_i-\sum^N_{j=0}X_j$ are the free concentration of receptors and ligands, with $R_T$, $L_T$ and $K_T$ the total number of receptors, ligands and kinase respectively. For notational simplicity, throughout the manuscript we assume that cell volume is fixed and hence do not distinguish between species number and concentration. In Fig. \ref{fig2}, we set $N=4$ and $m=2$.  More information about molecular species and notation can be found in Table. \ref{table1}.

\begin{table}
\centering
\caption{Definition of symbols shown in Fig. \ref{fig2}}
\label{table1}
\begin{tabular}{lc}
\hline
Symbol & Definition  \\
\hline\hline
$C_n$ & Agonist complex phosphorylated n times \\
$D_n$ & Non-agonist complex phosphorylated n times \\
$R$ & Receptor  \\
$K$ & Active kinase  \\
$K^*$ & Inactive kinase  \\
$K_T$ & Kinase \\
$\kappa$ & Ligand-receptor binding rate \\
$\phi$ & Complex phosphorylation rate  \\
$\alpha K$ & Complex phosphorylation rate at the final step \\
$b$ & Complex dephosphorylation rate  \\
$\sigma$ & Kinase phosphorylation rate  \\
$\epsilon$ & Kinase dephosphorylation rate  \\
\hline
\end{tabular}

\end{table}

In the adaptive sorting network,  both foreign and self ligands can bind a receptor and form the receptor-ligand complex, $X_0$,  which can undergo multiple rounds of phosphorlyation ($X_n$ goes to $X_{n+1}$) and dephosphorylation ($X_n$ goes to $X_{n-1}$).  The receptor-ligand complexes can disassociate  (at a rate $\tau_s ^{-1}$ for self ligands and $\tau_f^{-1}$  for foreign ligands). During this process, the phosphate groups are lost and and whole process reinitiates. Importantly, once a ligand is bound to a receptor, it is impossible for the biochemical machinery to distinguish between foreign and self ligands. The binding rate $\kappa$, the phosphorylation rate, $\phi$, and the dephosphorylation rate, $b$, inside the cell are the same for the foreign and self ligands and the only difference between foreign and self ligands are the lifetimes of their corresponding receptor-ligand complexes. For this reason, the decision to activate is based on the concentration of the total final products $C_N+D_N$ from both the foreign ligand ($C_N$) or self ligand ($D_N$).  

In the adaptive sorting network, in addition to the phosphorylation cascade, a negative feedback is used to modulate the phosphorylation and/or dephosphorylation rates \cite{franccois2008case,franccois2016case}. For example, in Fig. \ref{fig2} the last phosphorylation step, from $X_{N-1}$ to $X_N$, is modulated by the level of active kinase $K$, which itself is dependent on the concentration of the $m$-th intermediate concentration $X_m$ through phosphorylation. With this feedback, the output signal is independent of the ligand concentration and only replies on the value of $\tau$. This model reduces to a KPR cascade when the feedback is absent, $i.e.$ $\epsilon=0$ and $\alpha=\phi/K_T$.

In many treatments of KPR, especially in the context of T-cell discrimination, the dissociation of the receptor-ligand is often treated as an irreversible process ($\gamma=0$). Often, this is a good approximation since phosphatases can easily bind  free receptors and quickly remove phosphate groups from the receptors \cite{mckeithan1995kinetic}. For this reason, in most studies that seek to model T-cell discrimination, it is sensible to set $\gamma=0$.   Here, we assume this rate is finite and small ($\gamma\ll 1$).   The reason for this choice is that rather than focus purely on the biologically relevant regimes, the goal of this study to make a phase diagram of the performance of KPR-based TCR circuits in the speed, accuracy,  energy-consumption plane. Below, we show that taking $\gamma \neq 0$ is essential to constructing an accurate phase diagram and identifying a feasible operating region in the speed-energy-accuracy plane.

 In any thermodynamically consistent model, all reactions are reversible and it is important to consistently treat both the forward rate and backward rate for the formation and disassociation of a complex. Let $\gamma_{n,i}$ denote the rate at which a self  ($i=s$) or foreign ligand ($i=f$) can directly form a complex at $n-th$ step of the KPR cascade (see Fig. \ref{fig2}). In such a reaction, the first $n-1$ steps of the KPR cascade are bypassed resulting in lower accuracy. There are several natural choices for how to choose $\gamma_{n,i}$. One common choice in the literature is to assume that  $\gamma_{n,i}$ is independent of $n$ and given by $\gamma_{n,i}=\gamma/\tau_i$. However, with this choice never saturates the KPR  accuracy bound for an N-step cascade,  $\eta_{min}=\tau_s^N/\tau_f^N$,  especially when N is large (see Appendix). 

For this reason, in this work we choose a step-dependent rate,  $\gamma_{n,i}=\gamma^n/\tau_i$ $(i=s,f)$, for directly forming a complex $C_n$  and $D_n$  This functional form is a direct consequence of assuming that  there is a constant free energy difference $k_B T\mathrm{log}{\phi}/{\gamma b}$ per phosphorylation. Having a large $\gamma$ will result in a bypassing of the proofreading steps and a high error threshold for any KPR-based circuit\cite{hopfield1974kinetic}. At a biophysical level, a non-zero $\gamma$ models complicated microscopic processes that allow for the bypassing of the KPR cascade \cite{dushek2009role}.

\section*{ Defining Accuracy, Speed, and Dissipation}
Before analyzing the biochemical network outlined above, it is necessary to define accuracy, energy consumption, and speed for T-cell recognition in greater detail.

\subsection*{Accuracy} 

Recall, that a T-cell makes the decision to activate based on the total concentration of the full phosphorylated complexes $C_N+D_N$ from both the foreign ligand ($C_N$) and self ligand ($D_N$).  Ideally, T-cells are activated only in response to foreign ligands. Thus, following Hopfield \cite{hopfield1974kinetic} we can define the error rate $\eta$ as the ratio of  $C_N$ and $D_N$:
\begin{equation}
\eta=\frac{D_N}{C_N}.
\end{equation}
The concentrations of different components can be calculated by solving the deterministic equations  (\ref{eq1}) at steady state. In the immune recognition by T cells, it is important to achieve a small error rate $\sim 10^{-4}-10^{-6}$. For an irreversible $N$-step KPR process (i.e. $\gamma=0$), $\eta$ can reach a minimum value we dub the ``Hopfield limit''
 \begin{equation}
\eta_{min}= {\tau_s^N}/{\tau_f^N}.
\end{equation}
We define the accuracy as one minus the error rate, $1-\eta$.

\subsection*{Energy Consumption}
In any non-equilibrium steady state, detailed balance is broken and leading to the existence of net currents in the network\cite{landauer1961irreversibility, hill2012free}.
The chemical potential difference between the reactants and products can be written as
 \begin{equation}
\Delta\mu=k_BT\mathrm{ln}\frac{J_+}{J-}
\end{equation}
where $J_+, J_-$ are forward- and backward-reaction fluxes. The net current is $J=J_+-J_-$. The power dissipation is defined as \cite{hill2012free,qian2007phosphorylation}
\begin{equation}
W=k_BT J\mathrm{ln}\frac{J_+}{J-}
\end{equation}
For example, the power dissipation of the first-step phosphorylation process: \ce{$C_0$ <=>[\ce{$\phi$}][\ce{$b$}]$C_1$} can be calculated as:
\begin{equation}\label{eq5}
W=k_BT\left(\phi C_0-b C_1 \right)\mathrm{ln}\frac{\phi C_0}{b C_1 }
\end{equation}
This can be generalized to the full KPR cascade and adaptive network (see Appendix). Finally, we adapt the convention of non-equilibrium thermodynamics and use the phrases ``energy consumption'' and ``power dissipation'' interchangeably.
\begin{figure}
\centering
\includegraphics[width=0.45\textwidth]{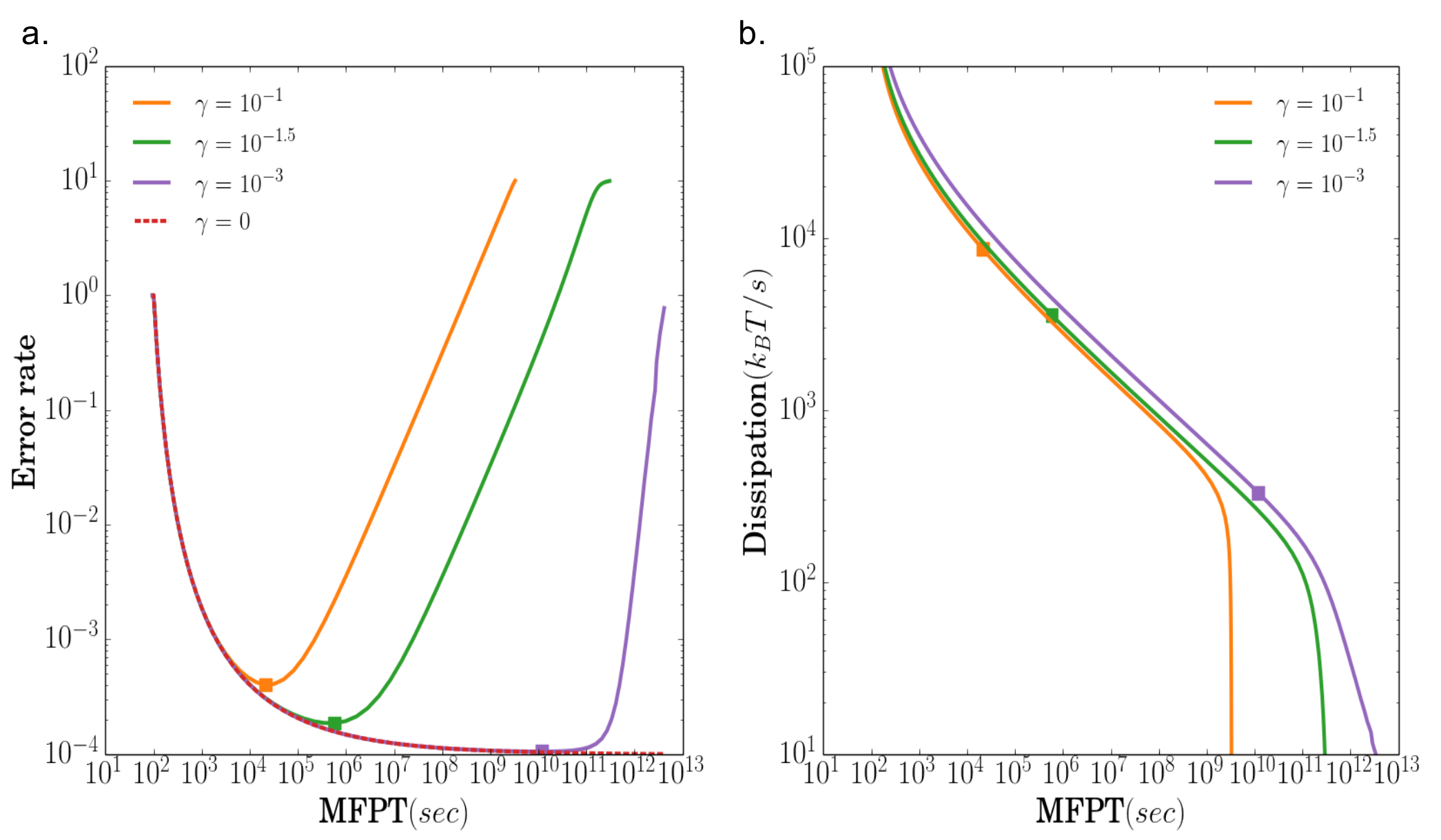}
\caption{{\bf Effect of changing $\gamma$ on the speed, accuracy, and energy consumption of a KPR based-circuit}. (a) Accuracy versus mean first-passage time (1/speed) and (b) dissipation (power consumption) versus mean first-passage time (1/speed) when the the phosphorylation rate $\phi$ is varied for different choices of  $\gamma=0, 0.1, 10^{-1.5}, 10^{-3}$. For all curves, $b/\phi=0.01$, $\tau_s=1$, and $\tau_f=10$. The squares indicate locations corresponding to parameters with a minimal error rate (see Fig \ref{fig2} and  Table. \ref{table1} for full definition of parameters).}
\label{fig3}
\end{figure}
\subsection*{Speed}
 The speed of decision-making process is related  to the mean first passage time(MFPT) of a stochastic process\cite{srivastava2015first}. The MFPT  is defined as average the time taken to produce one molecule of the final product $C_N$ from the foreign ligand $L_f$ . For example, at each time step, one molecule of the complex $C_3$ can be phosphorylated  at a rate $\phi$ to yield $C_4$, or can be dephosphorylated at a rate $b$ to get a molecule to $C_2$, or alternatively decay rate $\tau_f^{-1}$ to yield a free receptor $R$.  Microscopically, this can be viewed a stochastic process -- similar to a random walk-- and different realization of this process will take different amounts of time. The MFPT is taken as the average time it takes to complete to get from the starting point to the target. We use the mean MFPT to define the inverse of the decision speed.  Detailed calculation procedures can be found in Appendix and \cite{bel2009simplicity} 
 
Calculating speed in the adaptive sorting network is technically much more challenging than in KPR due to the non-linearity introduced by the additional feedback loop. To overcome this difficulty, we employ a linear-response approximation around the steady-state optimal point when calculating the speed. Such linear-response approximations are commonly employed in engineering (e.g. gain, bandwidth)  and have been adapted with great success to analyze biochemical circuits \cite{detwiler2000engineering}. In the linear-response regime of adaptive sorting, the MFPT can be calculated using methods analogous to KPR (in Appendix and \cite{van1992stochastic, mehta2012energetic}).

There are various methods to analyze the speed of KPR, the forward rate for a single step\cite{rao2015thermodynamics, sartori2013kinetic}, the gap between the first and second eigenvalue of the master equation \cite{ohzeki2015mathematical}, the inverse of the smallest eigenvalue of the master equation\cite{lahiri2016universal}, and also the MFPT\cite{murugan2012speed, bel2009simplicity, sharma2011distribution}. In this work, we measure the speed using the MFPT because it accurately reflects the speed of the circuit even in the presence of rare reactions that can bypass proofreading steps. We note that measures of speed based on eigenvalues of the master equations are accurate only for long Markov chains (i.e. $N\rightarrow \infty$) dominated by nearest-neighbor transitions \cite{kim1958mean}. The circuits considered here operate very far from these regimes and for this reason the MFPT is a more accurate measure of the speed of the proofreading process. 


\section*{Results}
We now analyze the speed-energy-accuracy tradeoff in KPR and adaptive-sorting circuits. One difficulty involved in identifying general principles are the large number of parameters whose choice can dramatically change the properties of the underlying circuit (see Table. \ref{table1}). For this reason, we will take a strategy based on randomly sampling these parameters in numerical simulations and looking for accessible regions in the energy-speed-accuracy plane. This spirit is similar to the one used to identify robustness in the adaptation circuit of bacterial chemotaxis \cite{barkai1997robustness, ma2009defining}. We begin by analyzing a KPR cascade where the feedback loop from the kinase $K$ in Fig. \ref{fig2} is turned off and then subsequently extend our analysis to the full adaptive sorting network.
\begin{figure}[t]
\centering
\includegraphics[width=0.45\textwidth]{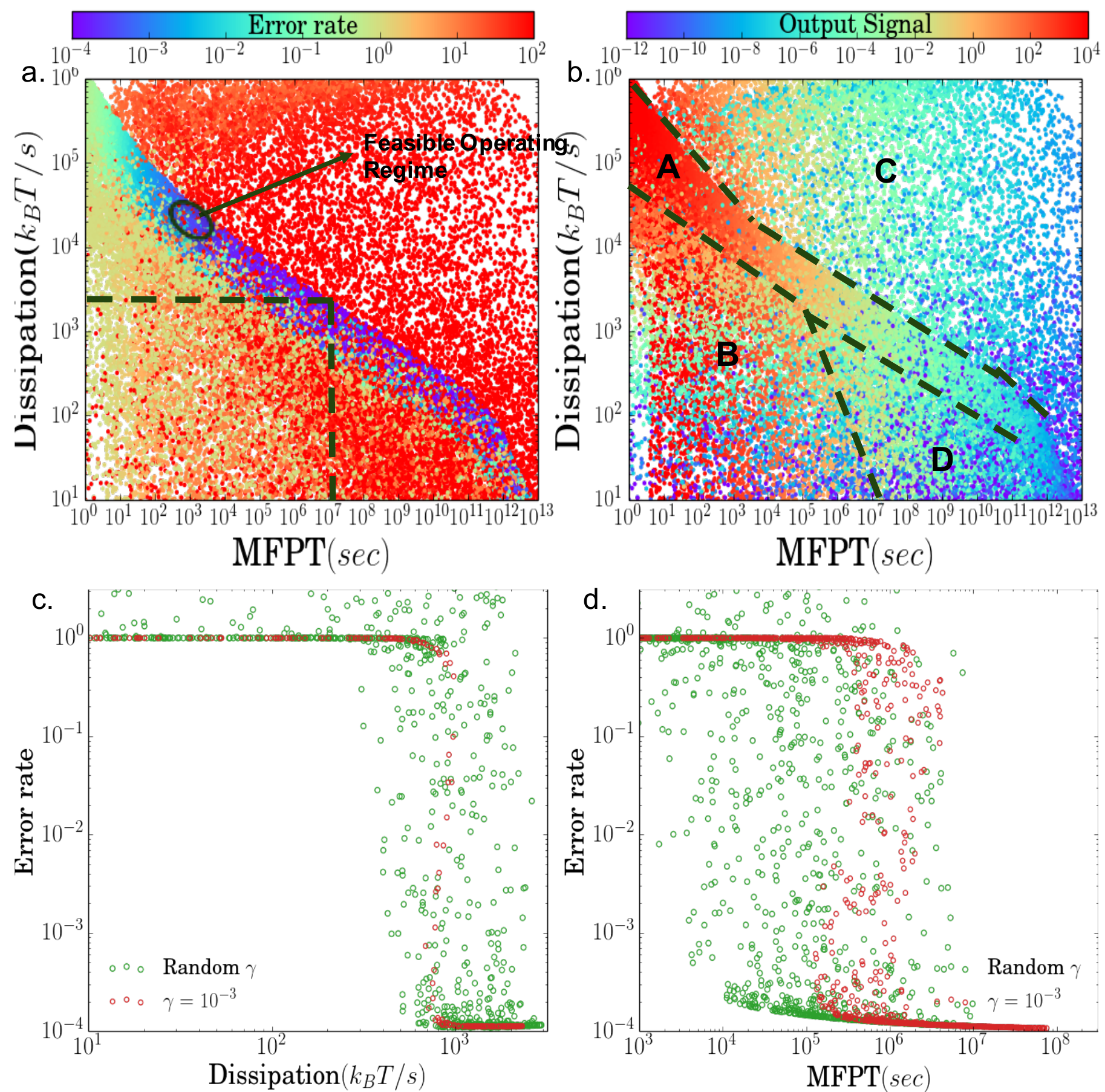}
\caption{Plots of the error rate and the magnitude of the output signal as a function of the  mean first-passage time (1/speed) and dissipation (power consumption) for randomly sampled parameters for the KPR circuit shown in Fig \ref{fig2}.  In generating these plots, we have kept the binding energy difference between self- and non-self ligands fixed by choosing $\tau_s/\tau_f=0.1$. (a)  Error rate; (b) Output signal: $C_4+D_4$; (c) Error rate versus dissipation for a $MFPT=10^7 s$ corresponding to the vertical dashed line in (a); (d) Error rate versus mean first passage time (inverse speed) at fixed dissipation rate$=10^{3.2}k_BT$ corresponding to the horizontal dashed line in (a).  The feasible region with high accuracy, high speed, low-dissipation, and a large output signal is labeled in (a). The behavior of circuit can be classified into four distinct regions labeled in A-D (see main text for more details).}
\label{fig4}
\end{figure}

\subsection*{Kinetic proofreading}

Some earlier theoretical works suggest that it is always possible to reduce the error of KPR-based mechanisms by waiting longer and/or consuming more energy\cite{savageau1979energy,murugan2012speed}. We find that this is not the case.   Our results show the error rate increases dramatically at extremely slow speeds/low dissipation when $\gamma$, the rate to directly form a complex that bypasses early KPR steps, has a nonzero value. This non-monotonic relationship between accuracy and speed was already noted as a possibility by Hopfield \cite{hopfield1974kinetic} and is consistent with a recent theoretical analysis of DNA replication and protein translation\cite{banerjee102608} and polymerization\cite{sartori2013kinetic}.

We studied the effects of varying $\gamma$ with numerical simulations shown  in Fig. \ref{fig3}. When $\gamma=0$, waiting longer always decreases the error rate. As shown in Fig. \ref{fig3}(a), the error rate $\eta$ monotonically decreases the with the MFPT (1/speed) and asymptotically reaches the Hopfield limit for an infinitely slow circuit: ${\tau_s^N}/{\tau_f^N}=10^{-4}$ for a circuit with $N=4$ phosphorylations.  In this high accuracy regime, a ligand must bind the receptor multiple times and transverse all $N$ steps of the phosphorylation cascade before reaching the final products $X_N$.  However, when $\gamma\neq 0$, for sufficiently long times, the probability to directly form a phosphorylated complex and bypass the initial kinetic proofreading steps becomes non-negligible. This leads to an increase in the error rate \cite{hopfield1974kinetic, murugan2012speed}.  Thus, increasing $\gamma$ drives a cross-over in the dynamic behavior of the biochemical circuit from a regime where waiting longer increases the accuracy to one where waiting longer decreases the accuracy.  

We also investigated the relationship between the speed of the circuit and power consumptions. Fig. \ref{fig3}(b) shows that over large parameter regime, the energy consumption and  MFPT (1/speed) exhibit an approximate power law (linear relationship on a log-log plot). This indicates that making a decision quickly always requires a a large amount of energy consumption. This approximate power-law relationship breaks down for extremely slow circuits.

In order to better understand the relationship between speed, accuracy, and energy consumption, we randomly sampled different combinations of the three parameters: $\phi$, $b$, $\gamma$ and calculated all three quantities(see Appendix for details). The results are shown in the Fig. \ref{fig4}(a). We also calculated the total output signal (the concentration of $C_N+D_N$) for each parameter set Fig. \ref{fig4}(b). {In defining this as the output signal of the KPR-circuit, we have assumed that the downstream machinery that reads out T-cell activation is sensitive to total concentrations of the output molecules. In the discussion below, we assume that if  the output signal is too small, it will be difficult for the molecular machinery downstream of the KPR machinery in T cells to activate a response.}

In both plots, each point corresponds to a different choice of the parameters. To better understand these plots, it is helpful to separate the parameters into four qualitatively distinct operating regimes (see  Fig. \ref{fig4}): (A) a high-accuracy regime, (B) a high-speed, low-dissipation, low-accuracy regime, (C) a high-dissipation, low-accuracy regime,  and (D) a low-dissipation, low-speed, low-accuracy regime. Region A is the discrimination regime, where the kinetic proofreading mechanism works;  Region B and D are close to the equilibrium state as the power dissipation is low and the error rate $\frac{D_N}{C_N}$ is close to 1; Region D is the ``anti-proofreading'' regime and there are large refluxes through the decay(discard) pathways \cite{Murugan2014, hartich2015nonequilibrium}.

One of the most dramatic features in Fig. \ref{fig4}(a) is the blue, high-accuracy region A. In Region A,  the error rate of the KPR cascade approaches its theoretically minimum possible value (i.e. the ``Hopfield Limit'') $\eta_{min}= \tau_s^N/\tau_f^N \approx 10^{-4}$ . This high accuracy region is realized when $\gamma\ll 1$, $b/\phi\ll 1$ and $\phi\ll \tau_s^{-1}$. These parameter regimes corresponds to the assumptions outlined by  Hopfield as being necessary for achieving high-accuracy proofreading \cite{hopfield1974kinetic}. Many choices of parameters in Region A achieve this high accuracy. 

However, as shown  Fig. \ref{fig4}(b) for many of these choices of parameters the magnitude of the output signal is quite small. This motivates defining a feasible
operating regime of the KPR regime as the choice of parameters with highest accuracy and a high output signal. This region is marked as the feasible operating regime in Fig. \ref{fig4}(a) (see discussion below).

In Region B, one can make a fast decision speed with minimal energy consumption, but the error rate is well above the Hopfield limit. Here, $b/\phi\ll1$ and $\phi\gtrsim\tau_s$. In this parameter regime, there is a steady-flux of empty receptors that are converted to the fully phosphorylated output complex. The MFPT is reduced but the system becomes insensitive to the difference between foreign and self-ligand binding times: the forward rate is so large that there is no time for the intermediate complexes to decay making it impossible to distinguish $\tau^{-1}_s$ and $\tau^{-1}_f$.Region C has the highest error rate. Here,  $\gamma\gtrsim1$, $b/\phi\gg1$ and $\phi\gtrsim\tau_s$. For such large values of $\gamma$, there is a continuous flux from free receptor directly to the fully-phosphorylated complex $C_N (D_N)$, with most output molecules bypassing the proofreading steps. In this region, $\gamma \ge \tau_f^{-1}, \tau_s^{-1}$ is much bigger than the binding times of ligands resulting in error rates that can be as large as $\eta=\tau^2_f/\tau^2_s=100$(see Appendix). In practice, for reasonable values of $\gamma$ (e.g. $\gamma \ll 1$), no biochemical networks operate in region C. Finally, in region D, speed decreases dramatically because of $\gamma\ll1$, $b/\phi\gg1$.

Fig. \ref{fig4}c and d show cross-sections of the error rate for a fixed speed and fixed dissipation rate respectively. These graphs were generated by selecting all parameters that lie along the vertical and horizontal dashed lines in Fig.  \ref{fig4}a. 
One of the most striking aspects of these plots is how dramatically the error rate decreases from the ``equilibrium value'' of $\tau_s/\tau_f=0.1$ to the theoretical maximum ``Hopfield limit'' $\left({\tau_s \over \tau_f}\right)^4=10^{-4}$ as a function of the dissipation rate and mean first-passage time. A similar plot for speed versus error rate was recently obtained by \cite{banerjee102608}.
Furthermore, the transition between these values become steeper and narrower as $\gamma$ is reduced. These plots suggest that for slow speeds (above $\sim 10^{-7} s^{-1}$) and low dissipation rates (below $\sim 10^3 k_BT/s$) there is maybe a dynamic phase transition in the KPR circuit when either the dissipation rate or speed is held fixed and other parameters are varied.

Murugan and collaborators have argued that KPR  has a natural mapping to microtubule growth, a system with a known dynamical phase transition between growth and shrinkage, and it has been argued that such a transition is also likely to be a generic feature of KPR \cite{murugan2012speed}.   However, unlike the systems analyzed by  \cite{murugan2012speed}, we consider a non-zero transition rate, $\gamma$, which leads to qualitatively different results. In particular,  our simulations show the existence of the low-fidelity region C in Fig. \ref{fig4} that arises when the mean first-passage time becomes comparable to the typical time it takes to ``bypass'' the KPR steps and directly form the complex $C_2$. 



\begin{figure*}
\centering
\includegraphics[width=1.0\textwidth]{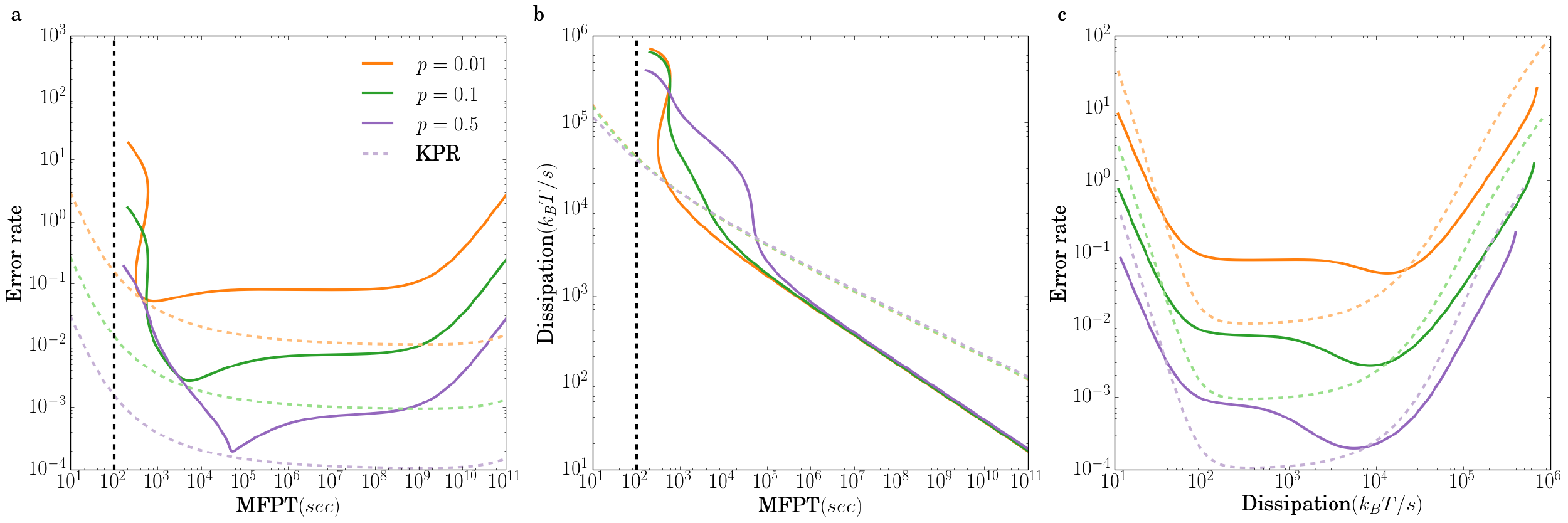}
\caption{\bf Effects of changing proportions of foreign ligand $p$ on the relations between Time, Dissipation and Error rate for adaptive sorting process.  }
$N=4$, $m=2$, $L_T = 2.0\times 10^4$, $L_f=pL_T$, $L_s=L-L_f$. $K_T=10^3$, $\gamma=10^{-3}$, $\sigma=2s^{-1}$, $\epsilon=1s^{-1}$, $\alpha K_T=3\times 10^{-1}s^{-1}$. The dashed lines are KPR results corresponding to different $p$. The black, vertical dashed line marks indicates the experimentally-measured time it takes T-cells to make decisions.
\label{fig5}
\end{figure*}
\subsection*{Extending our results to adaptive sorting}

In the preceding section, we have focused on the speed, accuracy, and dissipation trade-offs in a simple KPR cascade. Adaptive sorting is a very promising  extension of KPR relevant for understanding  T-cell activation in immune recognition \cite{franccois2008case,franccois2016case,lalanne2013principles,franccois2013phenotypic}. Adaptive sorting employs an additional negative feedback loop in the last step of the KPR cascade that ensures the output signal is independent of the number of ligands in the environment. This ability to perform ``absolute ligand discrimination'' is a key feature of adaptive sorting.  It accounts for how a T-cell can achieve high accuracy in  natural environmental conditions where the concentration of self-ligands is large and dwarfs the concentration of foreign ligands ($C_m\gg D_m$ and $C_m\gg 1$).  A natural question is to ask if there is any tradeoffs involved needed to achieve absolute ligand discrimination. One such tradeoff is antagonism, where increasing the concentration of foreign ligands actually degrades the response of the adaptive sorting circuit \cite{francois2016phenotypic}. We show here that there is another tradeoff between absolute ligand discrimination and the speed at which the T-cell receptor circuit can operate. 

Fig. \ref{fig5} shows error rate, mean first-passage time, and dissipation rate of the adaptive sorting and the KPR cascade analyzed above with regards to the tradeoffs between speed-accuracy and dissipations . The dissipation and error rate of the adaptive sorting model is comparable to a KPR cascade. However, from Fig. \ref{fig5}(a,b), it takes the adaptive sorting circuit much longer to achieve a similar error rate as a KPR.  For a very large input signal, the phosphorylation rate of the last step in the cascade is dramatically decreased, leading to dramatic decrease in speed because most complexes fall apart before reaching the final step of the cascade. Furthermore, notice that unlike KPR, the adaptive sorting circuit is unable to achieve even modest error rates for mean first passage times of 100s (vertical dashed lines in Fig. \ref{fig5}), corresponding to the experimentally observed time it takes T-cells to make the activation decision.  However, it is likely that other adaptive sorting circuit architectures can operate at faster speeds.

\section*{Discussion}

The immune system must quickly and accurately recognize foreign ligands.  To  carry out this task, the T-cells work out of equilibrium by actively consuming energy. This raises natural questions about the relationship between speed, accuracy, and energy consumption in two classes of biochemical networks that have been used to model immune recognition: a KPR-based network and a generalization of KPR, adaptive sorting. By numerically sampling parameter space, we found that the behavior of these networks exhibit  four different regimes, including a fast, high-accuracy regime at intermediate energy consumption which we call the feasible operating regime.

Our results also show that  waiting longer or consuming more energy does not necessarily translate into a higher accuracy. The underlying reason for this is that  we allow for a tiny (but) non-zero rate for bypassing the proofreading steps. While this parameter has no effect at short times, for very long times the error increases because the probability of bypassing the proofreading steps becomes significant even when absolute rates are small. Consist with this picture, recent works studying KPR in the context of DNA translation and polymerization have reached similar conclusions \cite{banerjee102608}. Moreover, the generality of this argument suggests that our conclusions should also hold for other, more complicated biochemical networks.

It has been argued that a KPR-based T-cell activation is likely to fail when the concentration of external ligands becomes large and one must instead consider an adaptive sorting based circuit \cite{franccois2008case,franccois2016case,lalanne2013principles,franccois2013phenotypic}. Unlike a simple KPR cascade, the adaptive sorting network can distinguish between foreign and self even for large ligand concentration, a property dubbed ``absolute ligand discrimination''. We have found that absolute ligand discrimination comes at a large cost in speed compared to a simple KPR-based circuit.  

We can compare our results for speed accuracy, and energy consumption to experiments. T-cells spend 1-5 mins to make the decision to activate \cite{feinerman2008quantitative}. A rough estimation of the error rate from experiment suggests cells can achieve error rates in the range$10^{-4}-10^{-6}$ or smaller, with the exact number depending on properties of ligands \cite{mckeithan1995kinetic,alon2006introduction}. {The energy expended by a T-cell to make the activation decision is hard to measure directly. However, estimates of the power consumption from glucose consumption suggest a typical cell uses about $10^9 \text{ATP}/s$ \cite{pollard2003cellular,milo2010bionumbers}. These numbers set strict experimentally-derived bounds for our model. }

For a circuit with $N=4$ phosphorylations, the minimum error rate achieved by both KPR and adaptive sorting is $10^{-4}$, on par with the experimental error rates. As shown in Fig.  \ref{fig5}, the KPR cascade can achieve close to this optimal accuracy in the experimentally observed decision time of $100s$. The power consumption of the circuit  is  $W\sim 1000 \text{ATP}/s$(where we have used the standard conversion $1\text{ATP}=14-15k_BT$ \cite{rosing1972value}),  just { one-one millionth} of the total energy budget of the cell. Moreover as shown in Appendix, increasing the number of steps in the phosphorylation cascade $N$ can significantly increase the accuracy of a KPR cascade with only modest decreases in the speed and the magnitude of the output signal. An adaptive sorting circuit can also reach the optimal error rate of  $~10^{-4}$ using approximately the same energy budget as a simple KPR cascade. However, the absolute ligand discrimination of adaptive sorting comes at a steep price in terms of speed.  For the biologically realistic $100s$ window for making immune recognition,  the KPR cascade achieves a respectable error rate between $10^{-3}$ and $10^{-4} $ whereas the adaptive sorting circuit is essentially non-functional. For this reason, it is also interesting to consider other mechanisms for balancing speed and accuracy \cite{kajita2017balancing}. 

Banerjee et al. point out biological systems prefer to optimize the speed rather than the accuracy \cite{banerjee102608}. Our simple model shows the high-accuracy proofreading regime is narrowly concentrated in the speed-energy consumption plane Fig. \ref{fig4} and the accuracy sharply decreases at the boundaries of this region. Within this high-accuracy region, the speed can change significantly -- the MFPT has a range from $10^2$ seconds to $10^{12}$ seconds --  but the accuracy does not fluctuate much. An analogous phenomenon was observed in the context of polymerization by \cite{banerjee102608}.

More generally,  the trade-off between speed, accuracy, and power consumption in realistic biochemical networks is still poorly understood.  Our results based on a simple model of immune decisions show that  thermodynamics places strict constraints on these non-equilibrium processes.  Energy consumption is required to maintain these non-equilibrium processes.  With extremely low energy consumption or slow speed, the decision signal will be ruined by thermal fluctuations. However, when operating in regimes with extremely large energy consumption or speed, subtle effects can suddenly transition circuits so that decisions are dominated by rare events that destroy accuracy. This suggests that great care is needed in both modeling and/or engineering KPR-based decision making circuits.

One of the most striking aspects of our simulations are the sudden transitions in accuracy as a function of the dissipation rate  (at fixed speed) or speed (at fixed dissipation). This transition seem to be indicative of an out-of-equilibrium dynamic phase transition. In the future, it will be interesting to further investigate this transition and see if it is possible to adopt analytic methods and fluctuation-type theorems to better understand its origins. Our work  also suggests that it is extremely difficult for adaptive sorting networks to simultaneously perform absolute ligand discrimination and operate quickly. An important area of future work is to better understand if this trade-off is fundamental or can be bypassed with more clever network architectures. Finally, it will be interesting to explore general networks and develop analytic techniques to further our understanding experimental operating regimes with regards to speed, accuracy, and power consumption.

\begin{acknowledgments}
This work was supported by NIH NIGMS MIRA grant number R35GM119461 and Simons grant in the Mathematical Modeling of Living Systems to PM.
\end{acknowledgments}
\onecolumngrid
\appendix

\section{ Definition of Model and parameter choices} 
\begin{figure}[tbhp!]
\centering
\includegraphics[width=0.7\linewidth]{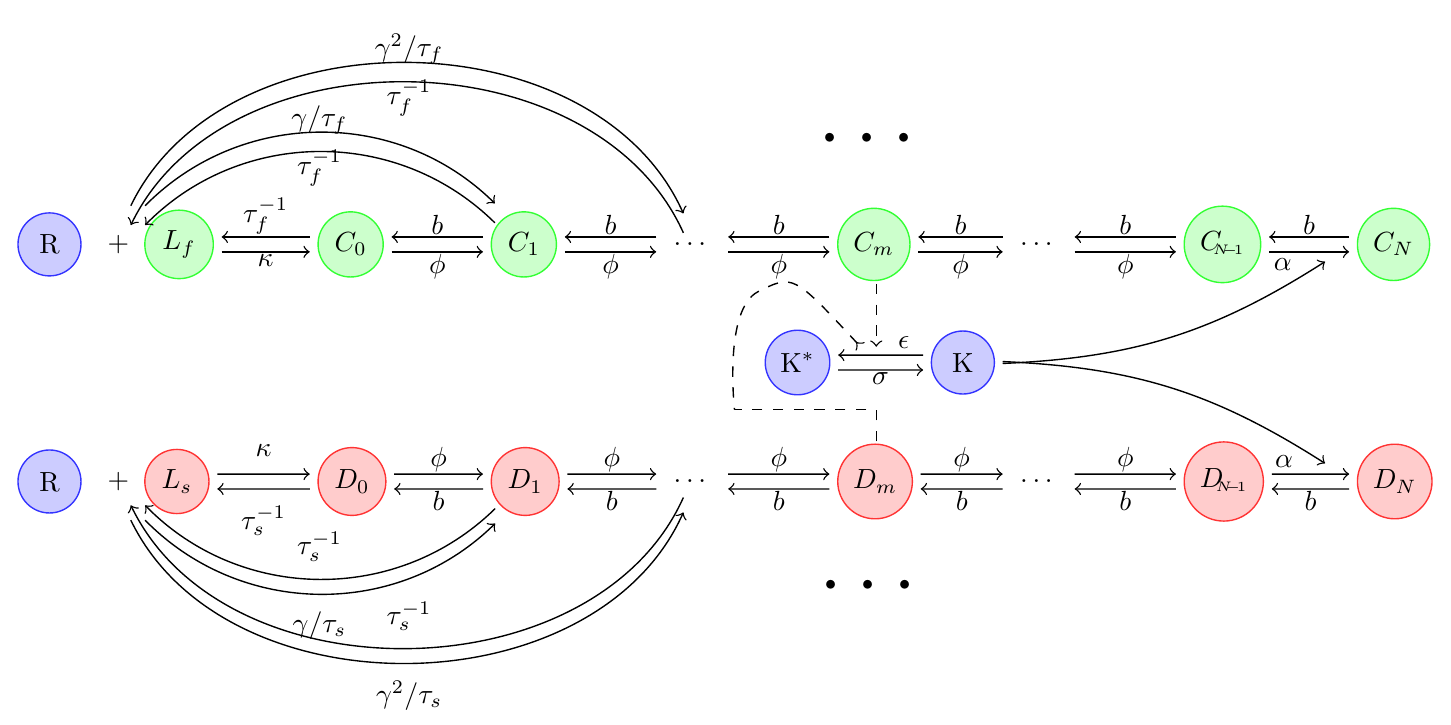}
\caption{Schematic overview of a general kinetic proofreading network.}\label{figs1}
\label{fig:SI1}
\end{figure}
A schematic of the model we are considering is shown in Fig. \ref{fig:SI1}. As described in the main text, we denote a receptor-ligand complex that has been phosphorlyated $n$ times by $X_n$ with $X=C$ for foreign ligands and $X=D$ for self ligands. Furthermore, we denote the maximum number of phosphorylations as N.
With this notation, using the law of mass action, we have
\begin{eqnarray}\label{eqSI1}
\dot{C_0}&=&\kappa RL_f-\lk\tau_f^{-1}+\phi \rk C_0+b C_1\nonumber\\
\dot{C_n}&=&\!\!\gamma^nRL_f/\tau_f \!+\!\phi C_{n-1}\!-\! (\phi\! +\!\tau_f^{-1}\!+\!b)C_{n}\!+\!bC_{n+1}\nonumber\\
\dot{C_N}&=&\gamma^N RL_f/\tau_f +\alpha K C_{N-1}-(b+\tau_f^{-1})C_N\nonumber\\
\dot{K}&=&-\epsilon K(C_m+D_m)+\sigma (K_T-K)\\
\dot{D_0}&=&\kappa RL_i-\lk\tau_s^{-1}+\phi \rk D_0+b D_1\nonumber\\
\dot{D_n}&=&\!\!\gamma^nRL_i/\tau_s\!+\!\phi D_{n-1}\!-\! (\phi\! +\!\tau_s^{-1}\!+\!b)D_{n}\!+\!bD_{n+1}\nonumber\\
\dot{D_N}&=&\gamma^N RL_s/\tau_s +\alpha K D_{N-1}-(b+\tau_s^{-1})D_N\nonumber
\end{eqnarray}
where $N>n>0$,  $R=R_T-\sum^N_{j=0}(C_j+D_j)\sim R_T$, $L_s=L^T_s-\sum^N_{j=0}D_j$ and $L_f=L^T_f-\sum^N_{j=0}C_j$. 
Typically, we set: $R_T=10^4$, $L^T_f=L^T_s=10^4$, $\kappa=300s^{-1}$, $\sigma=1 s^{-1}$, $\epsilon=2 s^{-1}$, $K_T=10^4$, $\alpha=3\times 10^{-4}$, $\gamma=10^{-3}$, $\tau_s=1s$ and $\tau_f=10s$.  Any deviations from this choice of parameter is explicitly noted. 

\subsection*{Accuracy}
At steady state, the error rate can be written as
\begin{equation}\label{eqSI2}
\eta=\frac{D_N}{C_N}
\end{equation}
In the presence of the kinase feedback $K=\frac{\sigma K_T}{\sigma +\epsilon (C_2+D_2)}$, the set of eqs. (\ref{eqSI1}) are no longer linear and but the steady-state solution can still be found easily using an iterative method. 
\subsection*{Energy Consumption}
The power dissipation is calculated based on the net flux and the chemical potential difference\cite{hill2012free,qian2007phosphorylation}.  We define the net flux $J_{\alpha,n}$ , $i\in[s, f]$ at $X_{n}\rightleftharpoons X_{n+1}$  in the main pathway. 
\begin{equation*}
J_{i,n}=\begin{cases}
\phi X_{n}-bX_{n+1}, \text{ for }0\le n<N-1\\
\alpha K X_{N-1}-b X_N, \text{ for } n=N
\end{cases}
\end{equation*}
Considering the flux conservation, the power dissipation $P_i$ can be written as
\begin{eqnarray*}
P_i&=&k_BTJ_{i,0}\mathrm{ln}\frac{\kappa R L^{free}_i}{\tau_i^{-1}X_0}+k_BT\sum^{N-2}_{n=0}J_{i,n}\mathrm{ln}\frac{\phi X_i}{bX_{i+1}}+k_BTJ_{i,N-1}\mathrm{ln}\frac{\alpha KX_{N-1}}{bX_{N}}\\
   &+&k_BT\sum^{N-2}_{n=0}(J_{i,n}-J_{i,n+1})\mathrm{ln}\frac{X_{n+1}}{\gamma^{n+1}RL_i^{free}}+k_BTJ_{i,N-1}\mathrm{ln}\frac{X_{N}}{\gamma^{N}RL_i^{free}}\\
&=&k_BJ_{i,0}\mathrm{ln}\frac{\kappa }{\tau_i^{-1}}+k_BT\sum^{N-2}_{n=0}J_{i,n}\mathrm{ln}\frac{\phi }{b\gamma}+k_BTJ_{i,N-1}\mathrm{ln}\frac{\alpha K}{b\gamma}\\   
\end{eqnarray*}
The total power dissipation is from the contribution of both foreign and self ligands: $P=P_s+P_f$.
\subsection*{Role of $\gamma$}

\begin{figure}[tbhp!]
\centering
\includegraphics[width=0.4\linewidth]{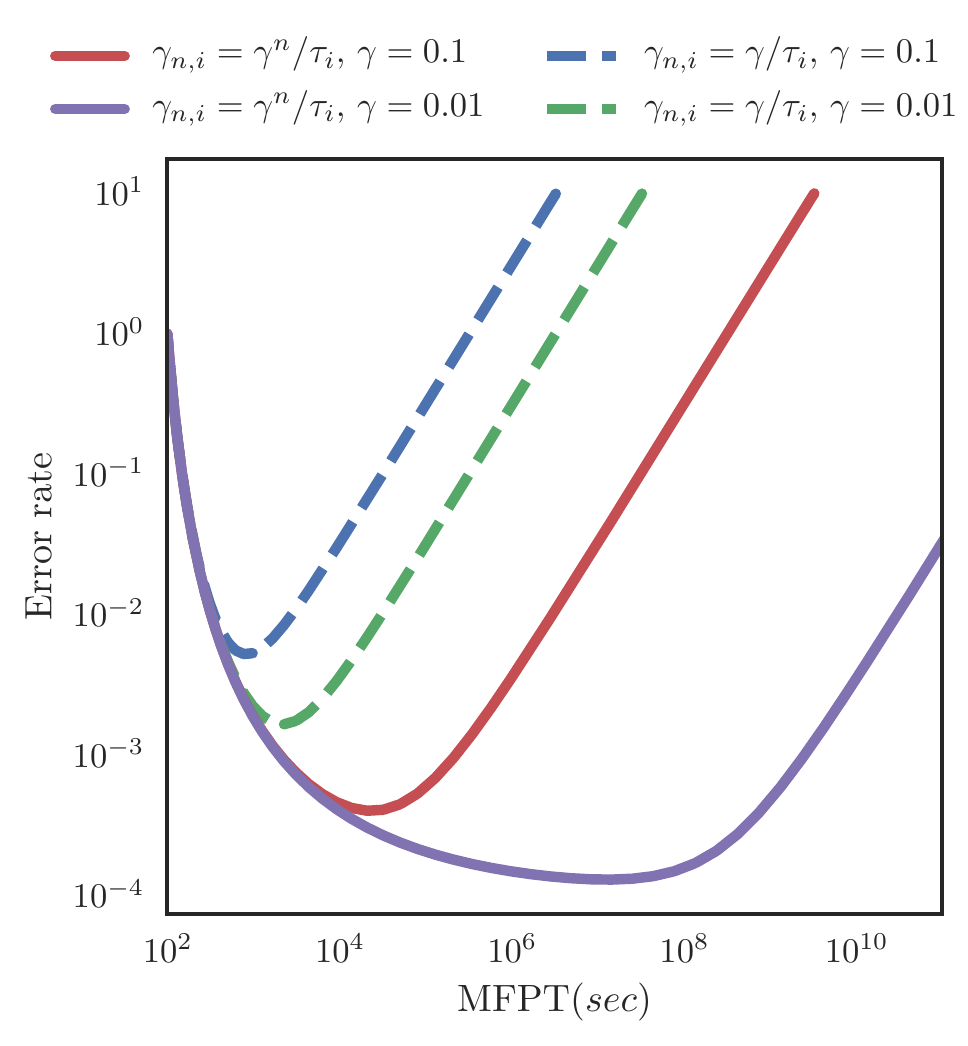}
\caption{Comparison between two different ways of parameterizing the rate $\gamma_{n,i}$: $\gamma_{n,i}=\gamma^n/\tau_i$ (solid lines) and $\gamma_{n,i}=\gamma/\tau_i$ (dashed lines).  If we assume all reversible decay rates are the same, it is impossible to saturate the theoretical ``Hopfield bound'' for accuracy, $\tau_s^N/\tau_s^N=10^{-4}$, when $\gamma_{n,i}$ is chosen according to the later scheme.}
\label{figs2}
\end{figure}

In KPR, the reversible decay rate is ignored as it has extremely small value. Let $\gamma_{n,i}$ denote the rate at which a self  ($i=s$) or foreign ligand ($i=f$) can directly form a complex at $n-th$ step of the KPR cascade (see Fig. 2 of main text). In such a reaction, the first $n-1$ steps of the KPR cascade are bypassed resulting in lower accuracy. There are several natural choices for how to choose $\gamma_{n,i}$. One common choice in the literature is to assume that  $\gamma_{n,i}$ is independent of $n$ and given by $\gamma_{n,i}=\gamma/\tau_i$. However, with this choice never saturates the KPR  accuracy bound for an N-step cascade,  $\eta_{min}=\tau_s^N/\tau_f^N$,  especially when N is large (see Fig \ref{figs2}). 

For this reason, in this work we choose a step-dependent rate,  $\gamma_{n,i}=\gamma^n/\tau_i$ $(i=s,f)$, for directly forming a complex $C_n$  and $D_n$  This functional form is a direct consequence of assuming that  there is a constant free energy difference $k_B T\mathrm{log}{\phi}/{\gamma b}$ per phosphorylation. Having a large $\gamma$ will result in a bypassing of the proofreading steps and a high error threshold for any KPR-based circuit.

 One choice o $\gamma^n/\tau_i$. There are two reasons for this form: 1. the production rate from ligands and receptors to $X_{n+1}$ should be smaller then the one to $X_n$ as one more phosphorylation step is involved. If not, it is hard for the KPR circuit to achieve the theoretical limit, $\tau_s^N/\tau_s^N$. 2. it is also natural to assume the energy consumption is the same for each phosphorylation step. 
 
The free energy difference between n$th$ and n+1$th$ phsphorlyation round can be calculated as:
\begin{eqnarray}
\Delta G_n&=&k_BT\mathrm{log}\left[ \frac{\kappa}{\tau_i^{-1}} \frac{\phi^{n+1}}{b^{n+1}}\frac{\tau_i^{-1}}{\gamma^{n+1}/\tau_i}\right]-k_BT\mathrm{log}\left[ \frac{\kappa}{\tau_i^{-1}} \frac{\phi^{n}}{b^{n}}\frac{\tau_i^{-1}}{\gamma^{n}/\tau_i}\right]\\
&=&k_BT\mathrm{log}\frac{\phi}{\gamma b}\nonumber
\end{eqnarray}
\subsection*{Speed}
The speed is defined by the mean first passage time(MFPT) for the foreign ligand. Here we mainly follow the procedures in  Ref. \cite{bel2009simplicity}.  The concentration vector is defined as $\mathbf{c}=\lK L_f, C_0, C_1,\dots, C_N ,p_f\rK$. An final 'dark' state is added because the response is only activated at the end and it can be treated as absorbing markov chain. Added this absorb state, it becomes an irreversible process, which is impossible to calculate the energy consumption. 
The transfer probability from $C_N$ to the 'dark' state is $W$(irreversible). We set $W=100s^{-1}$, a large value, which means the final step has little effect on MFPT. Without loss of generality, we begin with $N=4$ and $m=2$, which can be generalized other cases easily. The master equations $i.e.$ eqs. (\ref{eqSI1}) can be rewritten as $\dot{\mathbf{c}}=\mathbf{A}\mathbf{c}$ and 
\begin{equation}
\mathbf{A}\!=\!
\begin{bmatrix}
   - \kappa R \!-\sum_{j=1}^4\gamma^j/\tau_f& \frac{1}{\tau_f}& \frac{1}{\tau_f} & \frac{1}{\tau_f} & \frac{1}{\tau_f} & \frac{1}{\tau_f}& 0\\
 \kappa R                                &- \frac{1}{\tau_f}\!-\! \phi& b & 0& 0& 0& 0\\
\gamma/\tau_f         &   \phi           &- \phi-\frac{1}{\tau_f}-b & b& 0& 0& 0\\
\gamma^2/\tau_f    & 0                    &\phi&- \phi-\frac{1}{\tau_f}-b & b& 0& 0\\
\gamma^3/\tau_f     & 0       & 0             &\phi&- \alpha K-\frac{1}{\tau_f}-b & b& 0\\
\gamma^4/\tau_f    & 0       & 0             &0& \alpha K & -b-\frac{1}{\tau_f} -W& 0\\
0  & 0       & 0     &0        &0&  W& 0
\end{bmatrix}
\end{equation}

But eqs. (\ref{eqSI1}) are not linear. The first order perturbation approximation is adapted and we can linearize (with bar denoting average) to get $\mathbf{c}=\bar{\mathbf{c}}+\delta\mathbf{c}$.
\begin{equation}\label{eq5}
\delta\dot{\mathbf{c}}=\mathbf{A'}\delta\mathbf{c}, \quad \delta\mathbf{c}=\lK \delta L_f, \delta C_0, \delta C_1,\dots, \delta C_N, p_f \rK
\end{equation}
where $\mathbf{A'}$ is
\begin{equation}
\mathbf{A'}\!=\!
\begin{bmatrix}
   - \kappa R \!-\sum_{j=1}^4\gamma^j/\tau_f& \frac{1}{\tau_f}& \frac{1}{\tau_f} & \frac{1}{\tau_f} & \frac{1}{\tau_f} & \frac{1}{\tau_f}& 0\\
 \kappa R                                &- \frac{1}{\tau_f}\!-\! \phi& b & 0& 0& 0& 0\\
\gamma/\tau_f         &   \phi           &- \phi-\frac{1}{\tau_f}-b & b& 0& 0& 0\\
\gamma^2/\tau_f    & 0                    &\phi&- \phi-\frac{1}{\tau_f}-b & b& 0& 0\\
\gamma^3/\tau_f     & 0       & 0             &\phi+\frac{\alpha K \sigma C_3}{\epsilon+\sigma(C_2+D_2)}&- \alpha K-\frac{1}{\tau_f}-b & b& 0\\
\gamma^4/\tau_f    & 0       & 0             &-\frac{\alpha K \sigma C_3}{\epsilon+\sigma(C_2+D_2)}& \alpha K & -b-\frac{1}{\tau_f} -W& 0\\
0  & 0       & 0     &0        &0&  W& 0
\end{bmatrix}
\end{equation}
Applying the Laplace transform, $\delta\mathbf{C}(s)=\int_0^{\infty}\delta\mathbf{c}e^{-st}dt$, the master equations can be rewritten as:
\begin{equation}\label{eq7}
(s-\mathbf{A})\delta\mathbf{C}(s)=\delta\mathbf{c}(t=0)=\lK 1, \dots 0\rK^{T}
\end{equation}
The MFPT can be written:
\begin{equation}
T=\int_0^{\infty}t p_f(t)dt=-\frac{d}{ds}\int_0^{\infty} p_f(t)e^{-st}dt\vert_{s=0}=-W\frac{d\delta C_N(s)}{ds}\vert_{s=0}
\end{equation}
which can be calculated numerically. It should be notified that the concentration and probability have the same master equations but a different pre-factor. 
When choosing the initial condition $\lK 1, \dots 0\rK^{T}$, the pre-factor is set to be 1 and $\delta C_N(s)$ solved from eq. (\ref{eq7}) is exactly a probability distribution . 
\section{Simulation Details for Phase Diagram} 
In this figure, we run $10^6$ samples with random sets log uniformly chosen between $\gamma\in[10^{-4},10]$, $\phi\in[10^{-10},10^{10}]s^{-1}$, $b\in[10^{-15},10^{15}]s^{-1}$. 

It can be observed that a large amount of red points distributes over regimes C and D with $\eta\sim 100$. This is because of $\gamma\sim10$ and the inverse flux at the final step dominates.  In the extreme case: $b/\phi$ is very large, $C_0\simeq D_0$ will occupy most of products and free ligands $L_s, L_f$ have little concentration. 
$$\frac{L_s}{L_f}=\frac{D_0\tau^{-1}_s}{C_0\tau^{-1}_f}=\frac{\tau^{-1}_s}{\tau^{-1}_f}$$
As $\gamma^N/\tau_i$ dominates, $$\eta=\frac{D_N}{C_N}=\frac{L_s}{L_f}\frac{\gamma^N/\tau_s}{\gamma^N/\tau_f}=\frac{\tau_f^2}{\tau^2_s}=100$$
\section{Changing the number of phosphorylation steps}
Here, we show simulations for the KPR-cascade when we vary the maximum number of phosphorylation steps $N$.
\begin{figure}[h!]
\centering
\includegraphics[width=0.6\linewidth]{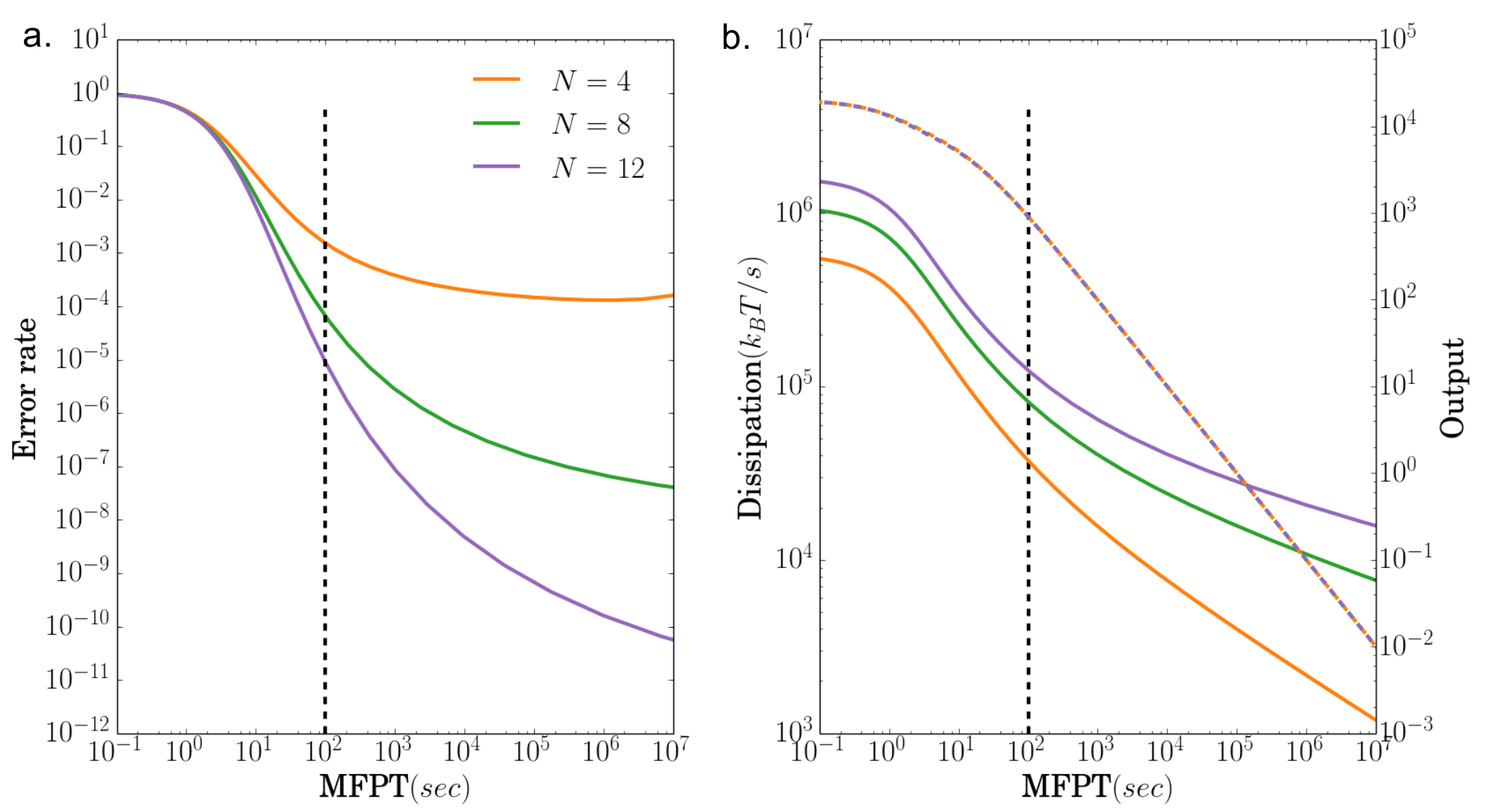}
\caption{Effects of the step size $N$ in KPR: $\tau_s=1s$, $\tau_f=10s$, we change $\phi$ but keep  $b/\phi=0.01$, $\gamma=10^{-3}$ fixed. The lines are for $N=4, 8$ and $12$. The vertical black dashed line is for $\text{time}=100s$. (a): relation between accuracy and speed;  (b): relation between dissipation(solid)/output(dashed) and speed.}
\label{fig:SI2}
\end{figure}

\bibliography{ref.bib}

\end{document}